# Micrometer thick soft magnetic films with magnetic moments restricted strictly in plane by negative magnetocrystalline anisotropy

Tianyong Ma, Juanying Jiao, Zhiwei Li*, Liang Qiao, Tao Wang\*, Fashen Li

Key Laboratory for Magnetism and Magnetic Materials of the Ministry of Education, Lanzhou University, Lanzhou 730000, China

Corresponding author: zweili@lzu.edu.cn and wtao@lzu.edu.cn

**Abstract**

Stripe domains or any other type domain structures with part of their magnetic moments deviating from the film plane, which usually occur above a certain film thickness, are known problems that limit their potential applications for soft magnetic thin films (SMTFs). In this work, we report the growth of micrometer thick c-axis oriented hcp-$Co_{84}Ir_{16}$ SMTFs with their magnetic moments restricted strictly in plane by negative magnetocrystalline anisotropy. Extensive characterizations have been performed on these films, which show that they exhibit very good soft magnetic properties even for our micrometer thick films. Moreover, the anisotropy properties and high-frequency properties were thoroughly investigated and our results show very promising properties of these SMTFs for future applications.

## 1. Introduction

In recent years, study of soft magnetic thin films (SMTFs) has become one of the hot topics in the field of magnetism because of their potential applications in high-frequency fields, such as miniature inductors, micro-transformers and noise suppressors [1-3]. In high-frequency applications, there are two basic demands for the SMTFs, which include small switching field and high permeability before the cut-off frequency. In order to reach these requirements, the magnetic materials must have both high saturation magnetization $M_s$ and low coercivity $H_c$.

Extensive studies have been devoted to Fe- and Co-based amorphous and nanocrystalline SMTFs with negligible magnetocrystalline anisotropy [4,5] and these SMTFs exhibit excellent static and high-frequency magnetic properties. However, in practical uses of magnetic devices, the SMTFs must have sufficient thickness (usually above micrometer) to obtain enough magnetic flux signal [6]. For the well-studied Fe- and Co-based amorphous and nanocrystalline SMTFs, a considerable perpendicular anisotropy will always occur due to defects and/or internal stress coming from the growth procedure of these SMTFs [7-9]. This perpendicular anisotropy will result in the appearance of a stripe-type domain structure when the thickness of these SMTFs exceeds a few hundred nanometers. Such a stripe domain structure will lead to two unfavorable consequences that hinders practical applications [10,11]: (1) a significant increase in coercivity, which leads to a deterioration of the soft magnetic properties; (2) the magnetic moments exhibit a large vertical component rather than strictly lie within the film plane. In practical uses, for example, the inductance and the quality factor of the miniature inductors will be significantly reduced and the loss will be extremely large when SMTFs with stripe domain structure were used. Therefore, preparation of thick enough SMTFs with its magnetic moments strictly lying parallel to the film plane is essential for them to have excellent soft magnetic properties and thus is crucial for practical applications of these SMTFs.

In our previous work, we have studied CoIr-based SMTFs with a strong negative magnetocrystalline anisotropy $K_u^{grain}$, which shows that the c-plane of the CoIr crystal is an easy magnetization plane and the c-axis is the corresponding hard axis. In other words, rotation of the magnetic moments from the c-plane to the c-axis direction must overcome an effective magnetic field $H_u^{grain} = \left| 2K_u^{grain} / M_s \right|$ of the order of ~8 kOe [12,13] in addition to the demagnetization field $4\pi M_s$. Therefore, we anticipate that the critical thickness for stripe type domain structure to appear shall be greatly enhanced if the CoIr SMTFs were grown with their c-plane parallel to the film plane. In this work, to verify our idea, we report the growth of oriented hcp-$Co_{84}Ir_{16}$ SMTFs up to micrometer thick. Our results show that these films with film thickness ranging from 100 nm to 1.12 μm all have excellent soft magnetic properties with small coercivity and, more importantly, with the magnetic moments of these films lie parallel to the film plane. The intrinsic magnetocrystalline anisotropy and total effective out-of-plane anisotropy have been investigated, and our high-frequency permeability measurements show very promising properties of these SMTFs for future practical applications.

## 2. Experiment

All our samples were prepared by DC perpendicular magnetron sputtering method with a layered structure of substrate/Ti/Au/$Co_{84}Ir_{16}$ (see inset of Fig. 1 (a)). Si wafer with (100) surface orientation was used as substrate, and the seed layer of Ti(8 nm) / Au(25 nm) was deposited first in order to induce the c-axis orientation of the $Co_{84}Ir_{16}$ soft magnetic layer. A variety of oriented $Co_{84}Ir_{16}$ films with different layer thicknesses ranging from 100 nm to 1120 nm were deposited using the same seed layer. In this work, for clarity, the samples were denoted as CoIr100, CoIr300, CoIr500, CoIr950, and CoIr1120 corresponding to samples with $Co_{84}Ir_{16}$ layer thicknesses of 100, 300, 500, 950, and 1120 nanometer. The seed layer was deposited using Ar as working gas with a pressure of 0.25 Pa. The $Co_{84}Ir_{16}$ soft magnetic layer was grown with an Ar pressure of 0.3 Pa by using Co target with five Ir chips symmetrically placed on the erosion race-track. To induce an in-plane uniaxial anisotropy, the substrate was inclined by 5º with a wedge during the sputtering process. The crystalline structure and cross section morphology were investigated by X-ray diffraction technique (XRD) and scanning electron microscope (SEM). The out-of-plane magnetization component in zero-field was detected through magnetic force microscopy (MFM). The static magnetic properties were measured with a vibrating sample magnetometer (VSM) and the dynamic magnetic properties were measured with an electron spin resonance spectrometer (ESR). The microwave permeability measurements were performed by a vector network analyzer (VNA) using the shorted microstrip method.

## 3. Results and discussion

The XRD patterns of the samples with indicated $Co_{84}Ir_{16}$ layer thicknesses measured with $\theta - 2\theta$ scan are shown in Fig. 1 (a). Only two diffraction peaks corresponding to Au (111) plane and $Co_{84}Ir_{16}$ (002) plane can be seen for all the samples. The schematic layer structure of the samples is shown in the inset of Fig. 1 (a). Similar to our previous work [13], the 8 nm Ti layer provides an amorphous surface that assists the oriented growth of the Au layer with (111) plane parallel to the thin film which was expected to induce the oriented growth of $Co_{84}Ir_{16}$ soft magnetic layer with its c-axis normal to the film surface. This was proved working very well by the only diffraction peak of the $Co_{84}Ir_{16}$ (002) plane for all our samples. To further verify the goodness of the c-axis alignment of the $Co_{84}Ir_{16}$ layer, rocking scans on the (002) peak were made for the CoIr100 and CoIr1120 samples as shown in Fig.1 (b). The full width at half maximum (FWHM) determined from Lorentz fits to the experimental data amount to 1.9º and 1.7º for the CoIr100 and CoIr1120 samples, respectively. The small values of FWHM reveal the good alignment of the $Co_{84}Ir_{16}$ layer [14]. Interestingly, the even smaller value for the CoIr1120 sample demonstrates that the misalignment which might come from defects or lattice mismatch can be further reduced for thicker samples. In Fig. 2 we present a typical cross-section image from SEM measurements for the 950 nm sample. The good alignment of the c-axis of the $Co_{84}Ir_{16}$ layer can be directly seen from the columnar-like crystalline structure.

The in-plane magnetic hysteresis loops of the oriented hcp-$Co_{84}Ir_{16}$ SMTFs are shown in Fig. 3 (a)-(e). These loops were measured with the applied magnetic field parallel or perpendicular to the easy axis within the film plane. For all samples, the hysteresis loops are totally different when measured in the two directions, which suggests that all films possess an in-plane uniaxial magnetic anisotropy. The uniaxial anisotropy induced by the oblique deposition technique can be explained with the framework of the so-called self-shadowing model [15]. According to this model, the region behind a growing crystallite is in the "shadow" of the crystallite and thus is prevented from continuous receiving metal vapor, so the crystallites will form a two-dimensional array of chains which leads to the uniaxial anisotropy of the SMTFs. The almost rectangular loops are observed along the easy axis and the remanence ratios are all above 0.92. The large remanence ratio suggests that nearly all the magnetic moments lie parallel to the film plane even for our micrometer thick sample. This is in sharp contrast to the Fe- and Co-based amorphous and nanocrystalline SMTFs in which strip domain structure will appear when the film thickness exceeds only a few hundred nanometers [7-9]. As shown in Fig.3 (f), the saturation magnetization $4\pi M_s$ of all the films are close to 14 KOe which agrees well with reported values in literature [16]. The coercivity $H_c$ determined from the easy axis loop decrease gradually from 35 Oe to 16 Oe with increasing film thickness. This decreasing tendency, which commonly exists in many magnetic thin films, can be well explained by the random anisotropy model for polycrystalline ferromagnetic films [17]. When the grain size is larger than the natural exchange length, the coercivity is inversely proportional to the grain size and can be described as $H_c \propto D^{-1}$ with nearly constant $4\pi Ms$ and nearly invariable in-plane uniaxial anisotropy field induced by the same condition.

The zero-field MFM images that were acquired in the as-deposited state at the film surface of sample CoIr950 and CoIr1120 are shown in Fig. 4. No obvious magnetic contrast can be seen in these zero-field MFM images, indicating the absence of stripe type domains and/or any other type of domain structures that involve with partial magnetic moments deviating from the others that lie parallel to the film plane. This matches well with the analysis from the hysteresis loops and is another evidence that most of the magnetic moments are restricted within the film plane. This is exactly the situation that we anticipated for the oriented hcp-$Co_{84}Ir_{16}$ SMTFs with a strong intrinsic negative magnetocrystalline anisotropy of the order of $\sim -10^6$ J/m$^3$. Together with the demagnetization field, this will provide an effective out-of-plane anisotropy of the order of $\sim 20$ kOe. On the other hand, the perpendicular anisotropy that leads to the stripe type domain structure is much smaller, $\sim 10^4$–$10^5$ J/m$^3$. For example, the reported perpendicular anisotropy is about 9.5 kJ/m$^3$ for NiFe films [18], and 11.3 kJ/m$^3$ for FeCoNbB amorphous film [19], where stripe domain structure appears above several hundred nanometers. Therefore, the magnetic moments in our SMTFs are completely compelled to lie within the film plane by the strong effective out-of-plane anisotropy field.

To quantify this effect and to get the value of the intrinsic magnetocrystalline anisotropy $K_u^{grain}$ and the total effective out-of-plane anisotropy field $H_\theta$ for our samples, we performed ESR measurements on all the films with a 9 GHz microwave magnetic field applied in-plane but perpendicular to the easy axis direction. The DC magnetic field is applied parallel to the film plane and forming an angle of $\varphi_H$ with the easy axis. The measured resonance fields $H_r$ are shown in Fig. 5 (a)-(e) as a function of $\varphi_H$ together with theoretical fits. It is well known that $H_r$ is smallest when the applied magnetic field is parallel to the easy axis and the largest when perpendicular. From published works [13,20], the relationship between the resonance field and the angle $\varphi_H$ can be expressed as:

$$\left(\frac{\omega}{\gamma}\right)^2 = \left[H_\theta + H_u\cos^2\varphi_0 + H_r\cos(\varphi_0 - \varphi_H)\right]\left[H_u\cos(2\varphi_0) + H_r\cos(\varphi_0 - \varphi_H)\right], \qquad (1)$$

where $\omega$ is the angular frequency, $\gamma$ is the gyromagnetic ratio, $\varphi_0$ is the equilibrium positions of the magnetization. The solid lines in Fig. 5 (a)-(e) are numerical fits to the experimental data using equation (1). The apparent twofold symmetry indicates a well defined in-plane uniaxial anisotropy in agreement with our VSM measurements [21]. The obtained total out-of-plane anisotropy field $H_\theta$ are shown in Fig. 5 (f) as a function of film thickness. As can be seen, $H_\theta$ decrease gradually from 27 kOe to 23 kOe with increasing film thickness. To get a clear understanding of this trend, we express the total out-of-plane anisotropy field as $H_\theta = 4\pi M_s + H_u^{grain} - H_p$, where $4\pi M_s + H_u^{grain}$ can be treated as constant. Then the decrease of $H_\theta$ is coming from $H_p$, the so called perpendicular anisotropy field in the favor of stripe domain structure. The corresponding perpendicular anisotropy constant can be determined as $K_p = H_p M_s / 2$ which we believe is originating from the columnar structure and any defects or stress of the as-grown film [22, 23]. Following the above definition, $4\pi M_s + H_u^{grain}$ is determined to be 27.5 kOe and the intrinsic magnetocrystalline anisotropy $K_u^{grain}$ was calculated to be -753 kJ/m$^3$ which agrees well with reported values [12,13]. Then the thickness dependence of the perpendicular anisotropy $K_p$ was calculated and shown in Fig. 5 (f). Although $K_p$ is much larger than that for the NiFe film and the FeCoNbB amorphous film where $K_p$ is in the order of 10 kJ/m$^3$, they are still much smaller than $\left| K_u^{grain} \right|$. Then, the net energy density gain $K_u^{grain} + K_p$ is still smaller when the magnetic moments lie parallel to the film plane even for the micrometer thick film. In other words, one can expect, the magnetic moments will be restricted within the film plane by the negative magnetocrystalline anisotropy provided $K_p \leq \left| K_u^{grain} \right|$.

To verify the potential applicability, we measured the microwave permeability spectra of our SMTFs by VNA. The frequency dependence of the permeability can be described by the following formula deduced from the Landau–Lifshitz–Gilbert (LLG) equation [24,25]:

$$\mu = 1 + 4\pi M_s \gamma \frac{\omega_1 + i\alpha\omega}{\omega_1\omega_2 - \omega^2 + i\alpha\omega(\omega_1 + \omega_2)}. \quad (3)$$

in which $\omega_1 = \gamma(4\pi M_s + H_u^{grain} + H_u) = \gamma(H_\theta + H_u)$, $\omega_2 = \gamma H_u$, where $\omega$ is the angular frequency, $\alpha$ is the damping constant and $H_u$ is the in-plane uniaxial anisotropy field. Considering the values of $M_s$ and $H_\theta$ have been obtained already, we fitted the experimental curves by equation (3) using $H_u$ and $\alpha$ as free fitting parameters as shown in Fig. 6 (a)-(e). The extracted values for $H_u$ are 115, 115, 120, 114, 110 Oe for our films in the order of increasing film thickness. Obviously, these values are approximately a constant which is consistent with the fact that they were induced by the same growth condition, namely the same incline angle ~ 5$^{\rm o}$ during the sputtering process [15]. One can see that all these films exhibit a similar behavior with a nearly constant real part before the gradually decreasing resonance frequency with increasing film thickness. The initial permeability matches well with the calculated values by equation $\mu_i = 1 + 4\pi M_s / H_u$, which is approximately 120 using the measured saturation magnetization $4\pi M_s$ and the extracted in-plane uniaxial anisotropy field $H_u$. The natural resonance frequency $f_r^{exp}$ corresponding to the maximum value in the imaginary part, as shown in Fig. 6 (f), follows the same trend of the calculated values by the extended Kittel's equation $f_r = \frac{\gamma}{2\pi}\sqrt{H_u H_\theta}$ which can be understood by the decrease of $H_\theta$ with increasing film thickness. The weak discrepancy of the experimental and calculated data may result from the revised Kittel's equation derived from the LLG equation by neglecting the damping effect [26].

The damping constant $\alpha$ derived from the fitting, which increase with film thickness, were plotted in Fig. 6 (f). It is well known that $\alpha$, which is an important parameter for high frequency applications [27], consists of two parts, namely, the intrinsic and extrinsic contributions. The intrinsic part is largely depends on fundamental properties such as Ms and spin-orbit coupling effect [28], and the extrinsic part is generally explained by the defect-induced two-magnon model and/or the local resonance model [29,30], both of which is associated with magnetic inhomogeneities within the material (e.g. due to anisotropy dispersion and surface or interface roughness). Here, the intrinsic part may not influence the damping constant since $M_s$ is almost a constant. Therefore, the extrinsic part must play an important role in increasing the damping constant. Apart from this, it was reported that eddy currents also affect the permeability spectrum and cause a broadening of the resonance peak for thicker films [31,32]. From the above considerations, we believe that both magnetic inhomogeneities and eddy currents are involved in the increase of the damping constant, especially for the micrometer thick films.

## 4. Conclusions

In summary, c-axis oriented hcp-$Co_{84}Ir_{16}$ magnetic films with strong negative magnetocrystalline anisotropy were fabricated in the range of 100 nm to 1.12 μm. All these films were shown to have high saturation magnetization and high rectangular ratio with low coercivity in the easy direction. More importantly, all these films have a high degree of c-axis orientation even up to micrometer thick. As a result, the stripe type domain structure that usually appears above a few hundred nanometers has been effectively suppressed so that the magnetic moments lie strictly within the film plane. Additional, the intrinsic negative magnetocrystalline anisotropy constant was determined to be $K_u^{grain}$ = -753 kJ/m³ and the permeability measurements indicate that all these films have very good high frequency properties. Therefore, our work provides an effective method to grow thick enough magnetic films with excellent soft magnetic characteristics for future practical applications.

## Acknowledgments

This work was supported by the National Natural Science Foundations of China (Nos. 11574122 and11204115) and the Fundamental Research Funds for the Central Universities (lzujbky-2017-k20).

## References

[1] X. Yang, J.Q. Wei, X. H. Li, L.Q. Gong, T. Wang, F.S. Li 2012 Thickness dependence of microwave magnetic properties in electrodeposited Fe‒Co soft magnetic films with in-plane anisotropy Physica B. **407** 555

[2] V. Korenivski 2000 GHz magnetic film inductors J. Magn. Magn. Mater. **215** 800

[3] M. Yamaguchi, Y. Miyazawa, K. Kaminishi, H. Kikuchi, S. Yabukami, K.I. Arai, T. Suzuki 2004 Soft magnetic

applications in the RF range J. Magn. Magn. Mater. **268** 170

[4] L. T. Hung, Nguyen N. Phuoc and C. K. Ong 2009 Angular dependence of dynamic magnetic properties and magnetization orientation distribution of thin films J. Appl. Phys. **106** 063907

[5] T. J. Klemmer, K.A. Ellis, L.H. Chen, B. van Dover and S. Jin 2000 Ultrahigh frequency permeability of sputtered Fe – Co – B thin films J. Appl. Phys. **87** 830

[6] V. Korenivski, R. B. van Dover 1997 Magnetic film inductors for radio frequency applications J. Appl. Phys. **82** 10

[7] Nissim Amos, Robert Fernandez, Rabee Ikkawi, Beomseop Lee, Andrey Lavrenov, Alexander Krichevsky, Dmitri Litvinov and Sakhrat Khizroev 2008 Magnetic force microscopy study of magnetic stripe domains in sputter deposited Permalloy thin films J. Appl. Phys. **103** 07E732

[8] A. B. Kashuba, V. L. Pokrovsky 1993 Stripe domain structures in a thin ferromagnetic film Phys. Rev. B. **48** 14

[9] Marco Coïsson, Federica Celegato, Elena Olivetti, Paola Tiberto, Franco Vinai and Marcello Baricco 2008 Stripe domains and spin reorientation transition in $Fe_{78}B_{13}Si_9$ thin films producedby rf sputtering J. Appl. Phys. **104** 033902

[10] Yuanfu Lou, Ge Yin, Fu Zheng, Feilong Luo, Jianmin Bai, Dongping Wu, Jiangwei Cao and Fulin Wei 2012 Analysis of magnetic anisotropy of FeCoAlON thin films by the domain structures Materials Letters **79** 45

[11] Nguyen N. Phuoc and C. K. Ong 2013 Thermal stability of high frequency properties of gradient-composition-sputtered FeCoHf films with and without stripe domains J. Appl. Phys. **114** 023901

[12] Atsushi Hashimotoa and Shin Saito, Migaku Takahashi 2006 A soft magnetic underlayer with negative uniaxial magnetocrystalline anisotropy for suppression of spike noise and wide adjacent track erasure in perpendicular recording media J. Appl. Phys. **99** 08Q907

[13] T. Wang, Y. Wang, G.G. Tan, F.S. Li, S. Ishio 2013 Microwave magnetic properties of the oriented CoIr soft magnetic film with negative magnetocrystalline anisotropy Physica B. **417** 24

[14] R. A. Morris, Y. Inaba, J.W. Harrell and G.B. Thompson 2010 Influence of underlayers on the c-axis distribution in $Co_{80}Pt_{20}$ thin films Thin Solid Films **518** 4970

[15] D. O. Smith, M.S. Cohen and G.P. Weiss 1960 Oblique-Incidence Anisotropy in Evaporated Permalloy Films J. App. Phys. **31** 1755

[16] D. Hasegawa, S. Nakasaka, M. Sato, T. Ogawa and M. Takahashi 2006 Magnetization Process of h.c.p.-CoIr Nanoparticles With Negative Uniaxial Magnetocrystalline Anisotropy IEEE Trans. Magn. **42** 2805

[17] G. Herzer 1990 GRAIN SIZE DEPENDENCE OF COERCIVITY AND PERMEABILITY IN NANOCRYSTALLINE FERROMAGNETS IEEE Trans. Magn. **26** 1397

[18] Jeffrey McCord, Burak Erkartal, Thomas von Hofe, Lorenz Kienle, Eckhard Quandt, Olga Roshchupkina and Jörg Grenzer 2013 Revisiting magnetic stripe domains—anisotropy gradient and stripe asymmetry J. Appl. Phys. **113**

[19] Youxing Yu, Youran Yang, Yijiao Shan, Xiaofang Bi 2012 Abnormal substrate temperature dependent out-of-plane anisotropy in FeCoNbB amorphous films Appl. Phys. Lett. **101** 232408

[20] Shujuan Yuan, Baojuan Kang, Liming Yu, Shixun Cao, Xinluo Zhao 2009 Increased ferromagnetic resonance linewidth and exchange anisotropy in NiFe/FeMn Bilayers J. Appl. Phys. **105** 063902

[21] S. M. Rezende, J. A. S. Moura, and F. M. de Aguiar 1994 Ferromagnetic resonance of Fe(111)thin films and Fe(111)/Cu(111) multilayers Phys. Rev. B. **49** 21

[22] Shan X. Wang,and Jongill Hong 1999 Magnetic and Microstructural Characterization of FeTaN High Saturation Materials for Recording Heads IEEE Trans. Magn. **35** 2

[23] R.A. Morris, Y. Inaba, J.W. Harrell and G.B. Thompson 2010 Influence of underlayers on the c-axis distribution in $Co_{80}Pt_{20}$ thin films Thin Solid Films. **518** 4970

[24] N.N. Phuoc, F. Xu, C.K. Ong 2009 Tuning magnetization dynamic properties of Fe–SiO2 multilayers by oblique deposition J. Appl. Phys. **105** 113926

[25] S. Ge, S. Yao, M. Yamaguchi, X. Yang, H. Zuo, T. Ishii, D. Zhou, and F.Li 2007 Microstructure and magnetism of FeCo–SiO2 nano-granular films for high frequency application J. Phys. D: Appl. Phys. **40** 3660

[26] Shujuan Yuan, Baojuan Kang, Liming Yu, Shixun Cao, Xinluo Zhao 2009 Increased ferromagnetic resonance linewidth and exchange anisotropy in NiFe/FeMn bilayers J. Appl. Phys. **105** 063902

[27] N. Fujita, N. Inaba, F. Kirino, S. Igarashi, K. Koike, H. Kato 2008 Damping constant of Co/Pt multilayer thin-film media J. Magn. Magn. Mater. **320** 3019

[28] V. Kambersky 1970 On the Landau-Lifshitz relaxation in ferromagnetic metals Can. J. Phys. B. **48** 2906

[29] R. Arias and D. L. Mills 1999 Extrinsic contributions to the ferromagnetic resonance response of ultrathin films Phys. Rev. B. **60** 7395

[30] C. Chappert, K. L. Dang, P. Beauvillain, H. Hurdequint, and D. Renard 1986 Ferromagnetic resonance studies of very thin cobalt films on a gold substrate Phys. Rev. B. **34** 3192

[31] J. Ben Youssef, N. Vukadinovic, D. Billet and M. Labrune 2004 Thickness-dependent magnetic excitations in Permalloy films with nonuniform magnetization Phys. Rev. B. **69** 174402

[32] N. Fujita, N. Inaba, F. Kirino, S. Igarashi, K. Koike and H. Kato 2008 Damping constant of Co/Pt multilayer thin-film media J. Magn. Magn. Mater. **320** 3019

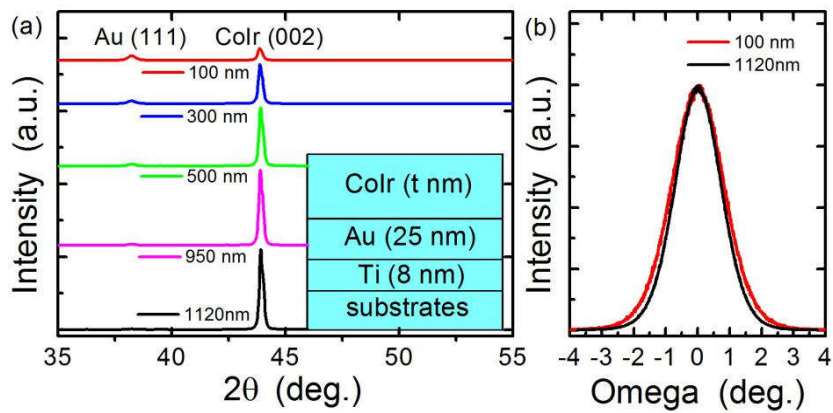

**Fig. 1** (a) XRD patterns of the oriented hcp-Co$_{84}$Ir$_{16}$ films with indicated layer thicknesses. The inset shows the schematic layer structure of the sample. (b) XRD rocking curves at the (002) peak of the 100 nm and 1120 nm films.

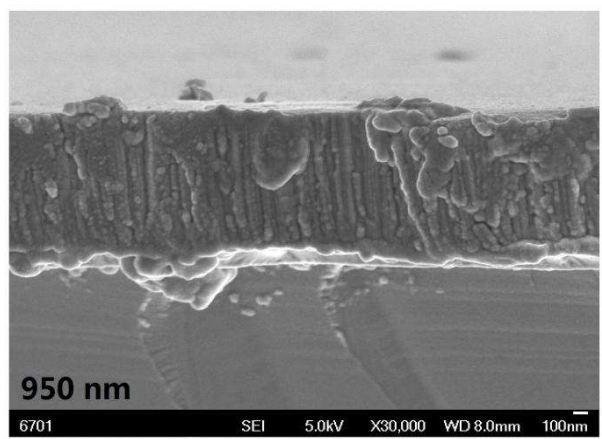

**Fig. 2** Typical SEM cross-section image of the oriented hcp-Co$_{84}$Ir$_{16}$ film of 950 nm thick.

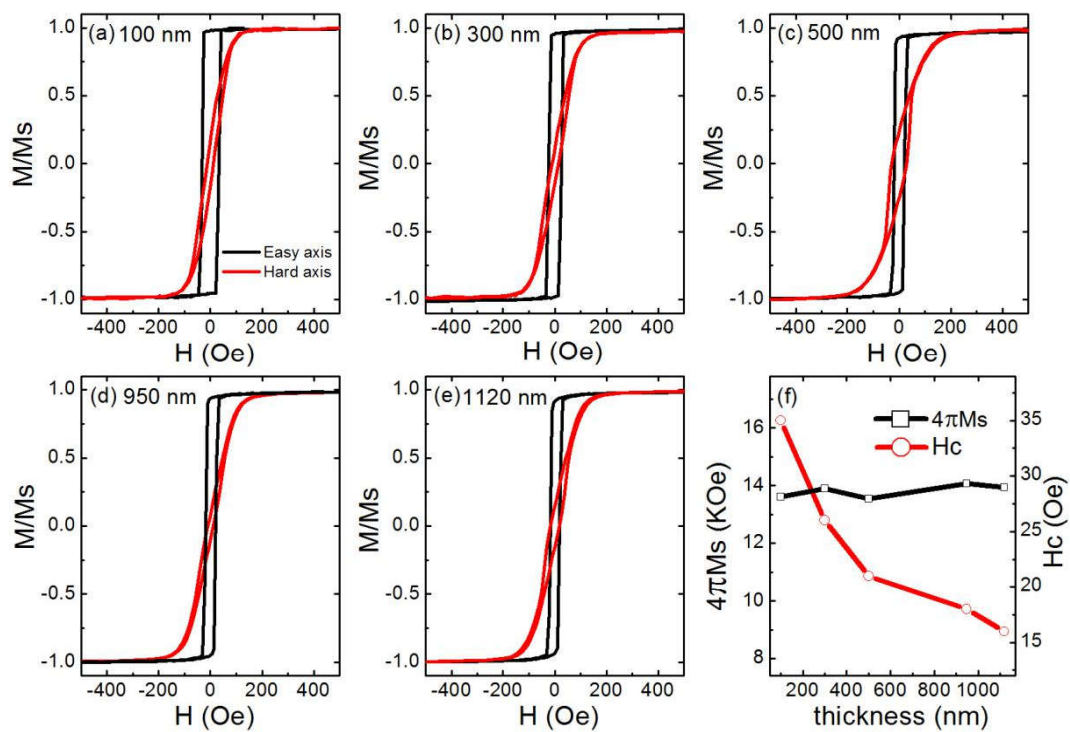

**Fig. 3** (a-e) In-plane magnetic hysteresis loops of the oriented hcp-Co$_{84}$Ir$_{16}$ films with the applied field parallel or perpendicular to the easy axis. (f) The thickness dependence of Hc determined from the easy axis loops and 4πMs for the oriented hcp-Co₈₄Ir₁₆ films.

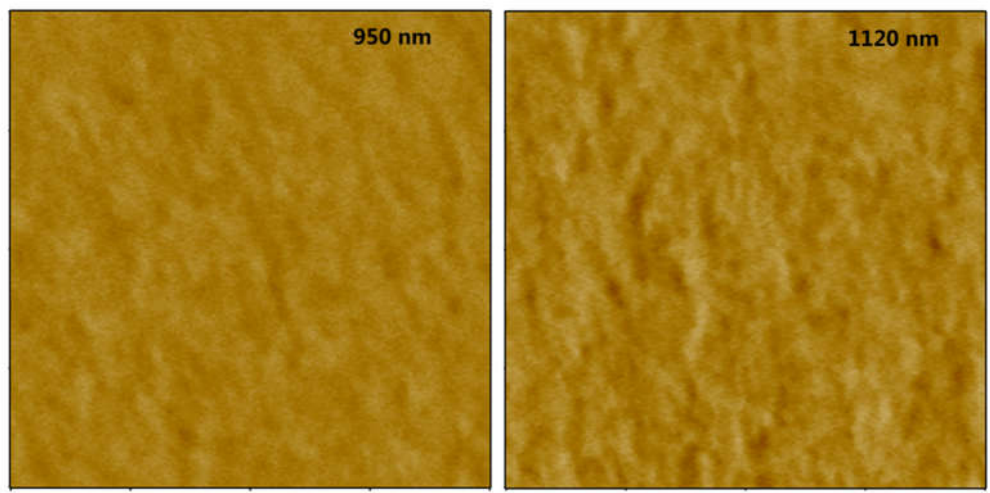

**Fig. 4** Zero-field MFM images of the oriented hcp-Co₈₄Ir₁₆ films with 950 nm and 1120 nm thicknesses. The image size are both 20 μm × 20 μm.

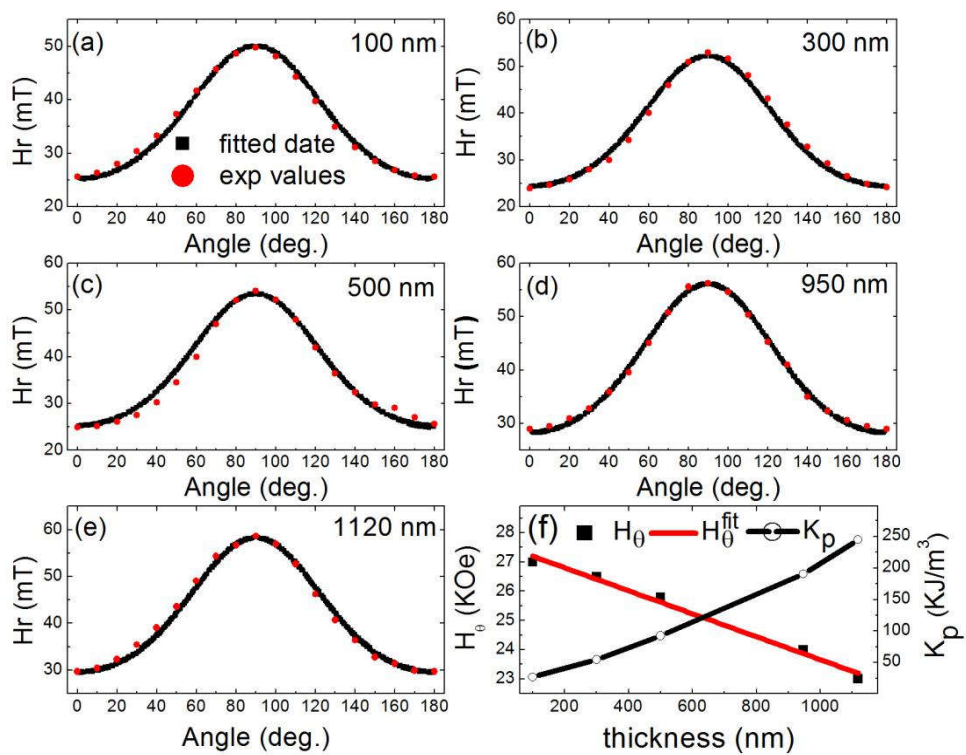

**Fig. 5** (a)-(e) Angle dependence of the resonance field (Hr) for the oriented hcp-Co₈₄Ir₁₆ films. The dots are experimental data and the curves are the fitted results. (f) The thickness dependence of $H_\theta$ and $K_p$. The red line is the fitting curve.

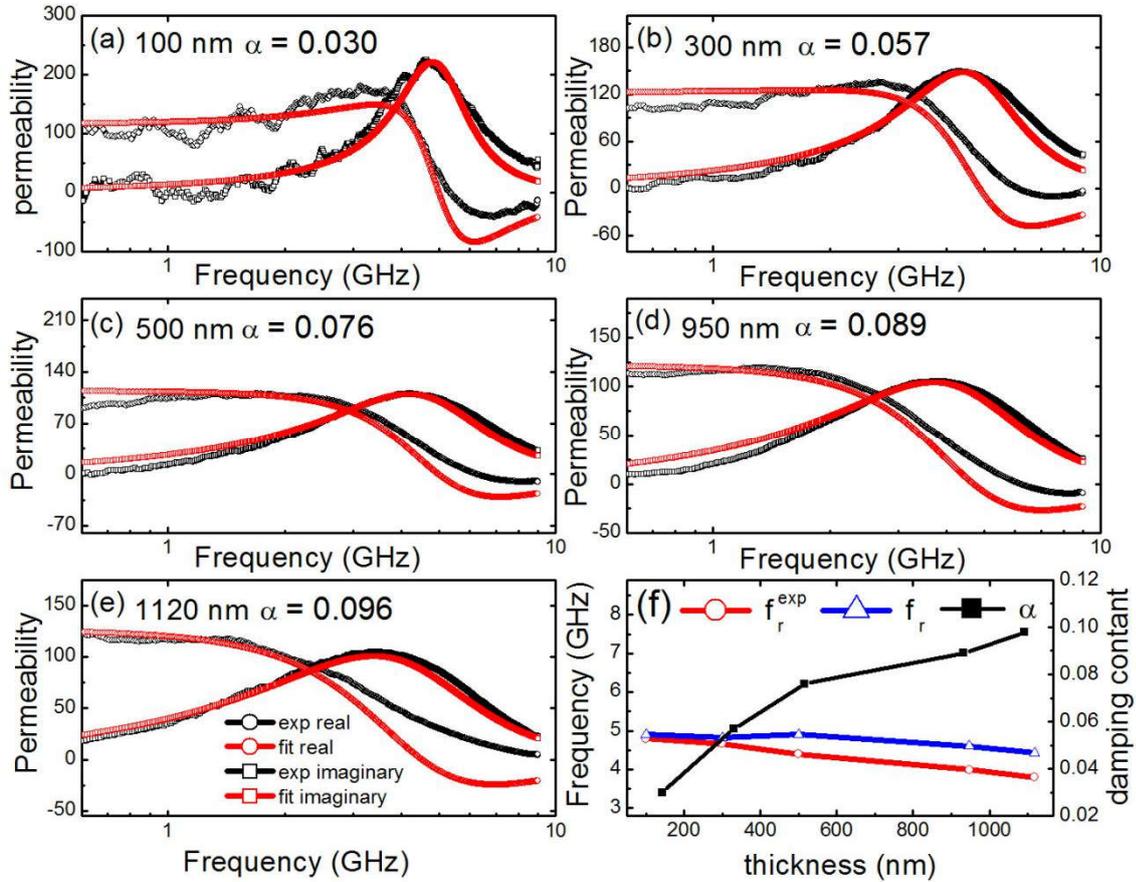

**Fig. 6** (a)-(e) Permeability spectra of the oriented hcp-$Co_{84}Ir_{16}$ films with indicated film thicknesses and the fitted damping constant α. (f) Thickness dependencies of the resonance frequency and the damping constant for the oriented $Co_{84}Ir_{16}$ films. $f_r$ is the calculated resonance frequency and $f_r^{exp}$ is that determined from the measured imaginary part of the permeability.